\documentclass[%
 reprint,
 amsmath,amssymb,
 aps,
]{revtex4-1}

\usepackage{graphicx}
\usepackage{dcolumn}
\usepackage{bm}

\begin{document}

\preprint{APS/123-QED}

\title{Time scales of epidemic spread and risk perception on adaptive networks}

\author{Li-Xin Zhong$^{1}$}\email{zlxxwj@163.com}
\author{Tian Qiu$^{2}$}
\author{Fei Ren$^{3}$}
\author{Ping-Ping Li$^{4}$}
\author{Bi-Hui Chen$^{1}$}
\affiliation{$^{1}$School of Journalism, Hangzhou Dianzi University, Hangzhou, 310018, China}
\affiliation{$^{2}$School of Information Engineering, Nanchang Hangkong University, Nanchang, 330063, China}
\affiliation{$^{3}$School of Business, East China University of Science and Technology, Shanghai, 200237, China}
\affiliation{$^{4}$School of Physics and Electronic Information, Wenzhou University, Wenzhou, 325027, China}

\date{\today}

\begin{abstract}
Incorporating dynamic contact networks and delayed awareness into a contagion model with memory, we study the spreading patterns of infectious diseases in connected populations. It is found that the spread of an infectious disease is not only related to the past exposures of an individual to the infected but also to the time scales of risk perception reflected in the social network adaptation. The epidemic threshold $p_{c}$ is found to decrease with the rise of the time scale parameter s and the memory length T, they satisfy the equation $p_{c} =\frac{1}{T}+ \frac{\omega T}{<k>a^s(1-e^{-\omega T^2/a^s})}$. Both the lifetime of the epidemic and the topological property of the evolved network are considered. The standard deviation $\sigma_{d}$ of the degree distribution increases with the rise of the absorbing time $t_{c}$, a power-law relation $\sigma_{d}=mt_{c}^\gamma$ is found.

\begin{description}
\item[Keywords]
Epidemic; Time scales; Risk perception; Rewiring

\item[PACS numbers]
89.75.Hc, 87.19.Xx, 89.75.Fb

\end{description}
\end{abstract}

\keywords{Epidemic; Time scales; Risk perception; Rewiring}

\maketitle


\section{\label{sec:level1}Introduction}
Over the last decade, the study of epidemiological processes on complex networks has become a great interest for scientists \cite{tang,parshani}. Among them, the susceptible-infected-susceptible (SIS) model and the susceptible-infected-refractory (SIR) model have been widely used in the study of epidemic dynamics on complex networks \cite{moore,zhou,pacheco,grabowski,fefferman}, such as the small world networks, the growing scalefree networks and the hierarchical networks \cite{kuperman,barthelemy,zheng}. It has been shown that the network geometry has a great impact on the epidemic threshold and the fluctuating endemic level \cite{colizza,watts}.

More recently, findings in the coupled dynamics of the epidemic status and the structural evolution have fueled extended discussions about the generalized contagion processes in the networked systems \cite{gross,shaw,son,volz}. In social contagion processes, an individual may get benefits or suffer from loss from existing relationships. For example, a close friend may help us overcome difficulties whereas an infectious individual may trigger influenza outbreaks in the neighborhood \cite{graser,buscarino}. When facing the threat of an epidemic, humans tend to avoid contact with the infected individuals. By rewiring a fraction of local connections, the social network evolves and the dynamics of the disease is influenced.

In the process of deactivating the links with potential hazards and creating new social ties, most coevolutionary epidemiological studies have made the additional assumption that the potential infectious hazards are known to all the individuals \cite{zhao,ni}. With complete information, it is easy for an individual to get rid of its disadvantage and get access to more social resources. However, because of the complexity of the real world, it is nearly impossible for us to have an accurate real-time information in all cases \cite{roca,zhong1,mosetti,funk,bagnoli,khan,xu}. For example, on Election Day, our knowledge of other people's opinions to a nominee only comes from the statistical data which were made public by the media a few days ago. In the stock market, the global information may not always be refreshed in time and the share holders can only make their judgment by the out-of-date or incomplete information. During an epidemic outbreak, as the outbreak of SARS in 2003, the information about who has been infected or who has recovered is not easy to be attained and our knowledge of the spread of epidemics usually lags behind the real-time information. Therefore, when modeling the coevolution of individual characteristics and socioeconomic networks, such timescale effects should not be disregarded.

In the study of the timescale effects between the epidemic information and the risk perception, the generalized contagion model introduced by P.S.Dodds et al \cite{dodds1} is especially useful to be studied upon. In the P.S.Dodds' epidemic model, the past exposures to the contagious influences are recorded in the memory of the agents. An individual's state is determined by his dose threshold. If the cumulative dose in the latest T time steps surpasses the dose threshold, the individual will become infected. Or else, he will be in the susceptible or recovery state. In such a generalized contagion model, the memory length T is one of the key factors in determining the fraction of infective individuals in the population in the final steady state. By introducing the rewiring process into the evolutionary steps, it is easy for us to know of how the delayed epidemic information will affect the time-dependent endemic levels and the topological properties of the evolved network.

Numerical simulations show that the delayed epidemic information may mislead the noninfected in the rewiring process, which will result in a decrease in the epidemic threshold and the prevalence of an infectious disease. One of the novel findings is that the evolved network structure is related to the time for an epidemic to reach steady state. Depending upon a mean-field approximation, both the degree distribution of the evolved network and the threshold condition are derived.

The plan of the paper is as follows. In Sec.II, the memory-based contagion model with timescales and adaptive networks is introduced. In Sec.III, we describe the implementation of the SIR dynamics and discuss the numerical simulation results about the spread of the infectious disease and the topological properties of the evolved network. In Sec. IV, theoretical analysis is given within the context of mean field theory. Our findings are summarized in Sec. V.

\section{\label{sec:level2}The model}
Following the work done in ref.\cite{dodds1}. In the present formulation of the memory-based contagion model, N individuals are initially located on a random regular (RRG) network with average degree $<k>$ and each individual is in one of the three states: S (susceptible), I (infected), or R (refractory). The SIR dynamics proceeds as follows. At each time step t, a randomly chosen individual i comes into contact with one of its immediate neighbors j and receives a dose d, which is kept in its memory recording the doses received in the latest $T$ time steps. The value of d is related to the state of j. If j is infected, i receives a positive dose $d>0$ with probability p; or else, $d=0$. Individual   changes its state according to the cumulative dose
\begin{equation}
D_{i}(t)=\sum_{t'=t-T+1}^{t} d_{i}(t'),
\end{equation}
i.e., if i is susceptible (infected) and $D_{i}(t)>d_{i}^{*}$ ($D_{i}(t)<d_{i}^{*}$), in which $d_{i}^{*}$ represents   i's dose threshold drawn randomly from an initially given distribution, i becomes infected (refractory).

To avoid being infected, a susceptible or refractory individual may rewire its local connections. The dynamics of the evolving network are as follows: with probability $\omega$, a susceptible or refractory individual can protect itself from being infected by cutting the existing links with the infected and establishing new links with other susceptible or refractory individuals. Double or self-connections are inhibited.

Time scales enter the coevolving dynamics through the rewiring process. There is a common look-up table recording the instant or delayed epidemic information. Keeping track of such information, a susceptible or refractory individual makes its decision on which link should be rewired. The timescale parameter s, an integer number between 0 and $\infty$, represents the delayed time step between the instant and the known epidemic information. For example, s=0 means that an individual knows about the instant epidemic information, that is, who has been infected and who has not been infected at the present time. With the instant information about each individual's state, a susceptible or refractory individual can make a right decision in the rewiring process. For $s\geq1$, the known epidemic information is s time step(s) delay in comparison with the instant epidemic information. A larger value of s corresponds to a slower reaction of an individual to the change of the states of other nodes and an S-S, S-R or R-R link may be cut by mistake.

In the coevolving process, the average degree $<k>$ is kept constant while the degree distribution P(k) evolves. The dispersion of degrees is measured by the standard deviation $\sigma_{d}$ of the degree distribution. The temporal evolution of the system can be captured by the ratio of the infected or the active S-I links. With the evolutionary dynamics, the system may finally reach one of the following two stable states: an absorbing state where no active link exists and an active stable state where the susceptible and the infected coexist with a fixed ratio.

\section{\label{sec:level3}Numerical simulations and discussions}
We have performed extensive numerical simulations to explore the timescale effects on the adaptive networks and the epidemic dynamics. Throughout the paper, dose sizes are lognormally distributed with mean one and variance 0.01, the dose cumulative time is set at T=10, the dose thresholds of all the nodes are the same as $d^{*}=1$, the average degree is $<k>=10$ and the population size is N=10000.

\begin{figure}
\includegraphics[width=3cm]{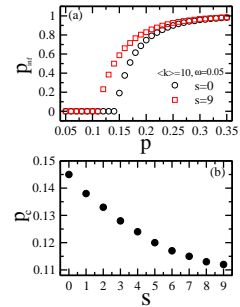}
\caption{\label{fig:epsart}(a)Average fraction of the infected in the final steady state as a function of the dose-receiving probability p for N=10000, $<k>=10$, $\omega=0.05$ and s=0 (circles), 9(squares), averaged over 10 runs. (b) The epidemic threshold   as a function of s.}
\end{figure}

Fig.1(a) shows the average fraction of infected nodes $p_{inf}$ as a function of dose-receiving probability p with the time scale parameter s=0 and 9. Each data point is obtained by averaging over 10 runs and after $10^4$ relaxation time steps for each run. The fraction of initially infected nodes and the rewiring probability are fixed at $p_{inf}^{0}=0.1$ and $\omega=0.05$. It is observed that, as p ranges from 0 to unity, there exist two thresholds, here we call them the epidemic threshold $p_{c}$ and the globally infected threshold $p'_{c}$ respectively. Below $p_{c}$ all the agents are in a disease-free state and above $p'_{c}$ the agents are all infected. Between $p_{c}$ and $p'_{c}$, the susceptible and the infected coexist and the fraction of the infected increases with the rise of p.

The epidemic threshold characterizes a transition between the disease free state and the epidemic prevalence in the steady state. The epidemic threshold $p_{c}$ decreases monotonically with the rise of s (Fig.1(b)). As s ranges from 0 to 9, $p_{c}$ shifts from 0.145 to 0.112 while $p'_{c}$ has little change. We have also found that $p_{c}$ and $p'_{c}$ are independent of the population size N.

\begin{figure}
\includegraphics[width=6cm]{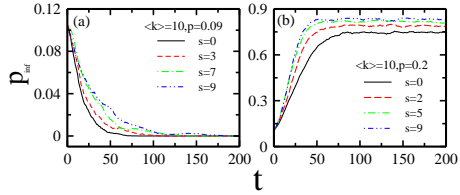}
\caption{\label{fig:epsart} The density of infected individuals versus time in an evolving network of N=10000, $<k>=10$, $\omega=0.05$ with (a) p=0.09, s=0(solid), 3(dash), 7(dash-dot), 9(dash-dot-dot) and (b) p=0.2, s=0(solid), 2(dash), 5(dash-dot), 9(dash-dot-dot).}
\end{figure}

To have a close eye of the temporal behaviors of epidemic spreading, in Fig.2 we give the density of infected individuals versus time for different p. For p=0.09, which corresponds to the region of $p<p_{c}$, the time to the disease-free state increases with the rise of s. For p=0.2, which corresponds to the region of $p>p_{c}$, increasing dose-receiving rate accelerates the dynamical evolution to the global infection. Such results indicate that the evolutionary time of the system to the equilibrium is closely related to the dose-receiving probability p and the timescale parameter s.

\begin{figure}
\includegraphics[width=3cm]{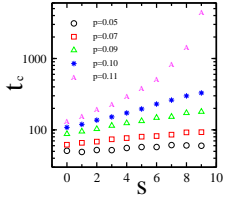}
\caption{\label{fig:epsart}The fixation time to the disease free state as a function of s for p=0.05(circles), 0.07(squares), 0.09(triangles), 0.10(stars), and 0.11(chars). All the other parameters are the same as that in Fig.2. Each data point is obtained by averaging over ten runs.}
\end{figure}

In Fig.3 the fixation time $t_{c}$ for the disease to go extinction is plotted against s for different p. It reveals that the functional relation between $t_{c}$ and s is determined by p. For small p, such as p=0.05, $t_{c}$ increases slowly with the rise of s. As p is close to the epidemic threshold, such as p=0.11, there is a sharp increase in $t_{c}$ as s increases.

\begin{figure}
\includegraphics[width=5cm]{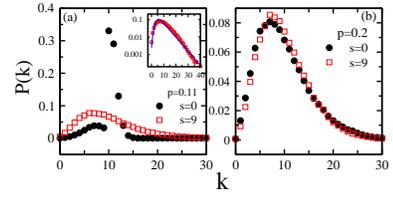}
\caption{\label{fig:epsart}The degree distribution of the evolved network for N=10000, $<k>=10$, $\omega=0.05$, (a) p=0.11 with s=0(circles), 9(squares) and (b) p=0.2 with s=0(circles), 9(squares), averaged over 10 runs. The inset gives the analytical result (solid line).}
\end{figure}

Rewiring not only leads to the change of the epidemic threshold but also the network structure. In Fig.4(a) and (b) we plot the degree distribution of the evolved network in the final steady state with s=0 and 9. Fig.4(a) shows the change of s has a great impact on the topological property of the evolved network within the range of $p<p_{c}$. For s=0, the evolution of the network leads to the increase of the average degree of some agents and the decrease of the average degree of other agents. The combination of a Possion-like distribution peaked at $<k>\sim7$ and an exponential-like distribution with a larger average degree forms the present curve. But for s=9, the evolution of the network leads to the occurrence of a lognormal degree distribution. As the  dose-receiving probability p is larger than $p_{c}$, Fig. 4(b) shows the change of s has only a minor impact on P(k), both of which are like lognormal distributions.

\begin{figure}
\includegraphics[width=3cm]{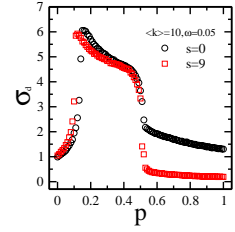}
\caption{\label{fig:epsart}The averaged standard deviation $\sigma_{d}$ of the degree distribution in the final steady state as a function of dose-receiving probability p for N=10000, $<k>=10$, $\omega=0.05$, and s=0(circles), 9(squares), averaged over 10 runs.}
\end{figure}

Fig. 5 shows the averaged standard deviation $\sigma_{d}$ of the degree distribution in the final steady state as a function of dose-receiving probability p for s=0 and 9. As p ranges from 0 to unity, $\sigma_{d}$ firstly increases with the rise of p and reaches a maximum value $\sigma_{max}$ near the epidemic threshold $p_{c}$. Then, $\sigma_{d}$ decreases slowly with the rise of p. As p is near the globally infected threshold $p'_{c}$, $\sigma_{d}$ has a sharp drop, which results from the limited evolutionary time for all the individuals to get infected because of the high infection rate. The change of s leads to the change of $\sigma_{d}$ within the whole range.

\begin{figure}
\includegraphics[width=5cm]{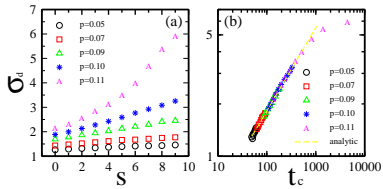}
\caption{\label{fig:epsart}The standard deviation $\sigma_{d}$ of the degree distribution as a function of (a) s and (b) $t_{c}$ with N=10000, $<k>=10$, p=0.05(circles), 0.07(squares), 0.09(triangles), 0.10(stars), and 0.11(chars), averaged over 100 runs. The dashed line satisfies the equation $\sigma_{d}=0.2t_{c}^{0.48}$.}
\end{figure}

To find the time scale effect on the evolved network structure, we give the standard deviation $\sigma_{d}$ of the degree distribution as a function of s in Fig.6(a). It shows that the functional relation between $\sigma_{d}$ and s is similar to that between $t_{c}$ and s (Fig.3). As the standard deviation $\sigma_{d}$ is plotted against $t_{c}$ in Fig. 6(b), it is observed that all the data have collapsed into a single line. When we perform a fit to the results in Fig.6(b), a power-law function $\sigma_{d}=mt_{c}^\gamma$,where $m\sim0.2$ and $\gamma\sim0.48$, is found. Such a result reveals that the change of $\sigma_{d}$ in the present model is determined by the evolutionary time for the system to reach equilibrium.

\section{\label{sec:level4}Analytical calculations}
\subsection{\label{subsec:levelA}evolutionary time and degree distribution}
In the present model, because of the switch process, the evolved network structure has significant alterations in comparison with the initial network connectivity. Using mean field analysis, in the following, we calculate the degree distribution of the evolved network with a fixed number of nodes and edges.

Firstly consider the degree distribution of the susceptible and the infected respectively irrespective of the infection process. Because the susceptible individuals protect themselves by reconnecting a fraction of S-I links into S-S links, the average degree $<k_{inf}>$ of the infected decreases while the average degree $<k_{sus}>$ of the susceptible increases with time. The random cut of S-I links forms a random graph of the infected, in which there is a Possion degree distribution with mean $<k_{inf}>$. But for the susceptible, the increase of the average link is just like that in the growing network without attachment preference introduced in ref.\cite{barabasi}, the degree distribution of them will evolve into an exponential distribution with mean $<k_{sus}>$.

Next, the epidemic is incorporated into the rewiring process. Because of the infection or recovery, the above independent cutting or liking process should take place in each individual. The degree of a node can be written as $k = k_{1} + k_{2}$, in which $k_{1}$ results from the random link and $k_{2}$ results from the growing link.

Considering the above two attachment methods, we can calculate the mixed degree distribution in the rewiring dynamics independent of the infectious process. Just as that in Ref.\cite{newman}, the random graph is attained as follows: start with n nodes. At each time step, two randomly chosen nodes are linked together with probability $\xi=\rho(1-q)$, where $\rho$ is the ratio of active links in the population and q is the growing attachment probability. The probability that a vertex has $k_{1}$ edges satisfies the equation
\begin{equation}
P_{k_{1}}=
\left(
\begin{array}{cc}
 n  \\
 k_{1}
\end{array}
\right)
\xi^{k_{1}}(1-\xi)^{n-k_{1}}.
\end{equation}
After a long time, the degree distribution of the evolved network follows a Poisson distribution
\begin{equation}
P(k_{1})\sim\frac{{e^{-\xi<k>}(\xi<k>)^{k_{1}}}}{k_{1}!}.
\end{equation}

In the growing model, a new node is linked to any of the nodes already existing in the system with the same probability \cite{barabasi}
\begin{equation}
\Omega=\frac{1}{m_{0}+t-1},
\end{equation}
where $m_{0}$ is the initial number of nodes in the system. With the time evolution, the degree distribution becomes a Possion distribution
\begin{equation}
P(k_{2})\sim e^{-\beta k_{2}},
\end{equation}
where $\beta=\frac{1}{q\rho<k>}$. Combining the above two mechanisms, the degree of a node in the present model can be written as $k=k_{1}+k_{2}$ and the degree distribution of the evolved network can be written as
\begin{equation}
P(k)\sim\sum_{k_{2}=0}^k\frac{[(1-q)\rho <k>]^{k-k_{2}}}{(k-k_{2})!}e^{-(1-q)\rho<k>}e^{-\frac{k_{2}}{q\rho<k>}},
\end{equation}
and the standard deviation of it becomes \cite{zhong2}
\begin{equation}
\sigma_{d}=\sqrt{(q\rho<k>)^{2}+(1-q)\rho<k>}.
\end{equation}

For a known $\sigma_{d}$, the growing attachment probability can be written as
\begin{equation}
q=\frac{1+\sqrt{1-4\rho<k>+4\sigma_{d}^2}}{2\rho<k>}.
\end{equation}
For $\rho=1$ and $<k>=10$, we obtain
\begin{equation}
q=\frac{1+\sqrt{4\sigma_{d}^2-39}}{20}.
\end{equation}

From simulation results in Fig.5 we observe that, near the threshold point p=0.11 for s=9, the standard deviation of the degree distribution is $\sigma_{d}=5.8$. From equation (9) we get q=0.54. Substituting q=0.54, $\rho=1$ and $<k>=10$ into equation (6), we get
\begin{equation}
P(k)\sim\sum_{k_{2}=0}^k\frac{(4.6)^{k-k_{2}}}{(k-k_{2})!e^{-(\frac{k_{2}}{5.4}+4.6)}}.
\end{equation}
Considering the normalizing condition, from Fig.4(a) we observe that the analytical calculations fit the simulation data well.

\subsection{\label{subsec:levelB}epidemic threshold}
The epidemic threshold is the critical infection probability $p_{c}$ above which a stable epidemic occurs. In the present model, $p_{c}$ is found to be related to the rewiring process which is determined by the delayed epidemic information. In the following, to derive the epidemic threshold, we assume that all the individuals have the same receiving dose d=1 or 0, dose threshold $d^{*}=1$ and memory size $T>1$.

First, consider the epidemic process without rewiring. At a given time, the contagious entity in the population consists of those who have had at least an S-I contact during the preceding T time steps. In the well-mixed case, the fraction of the infected satisfies the equation \cite{dodds2}
\begin{equation}
\varphi_{inf}=1-(1-p\varphi_{inf})^T.
\end{equation}
In the limiting case $\varphi_{inf}\to0$, the epidemic threshold is obtained
\begin{equation}
p_{c}=\frac{1}{T}.
\end{equation}
On a random graph, the epidemic threshold should also be related to the mean degree $<k>$[2]. To make the equation satisfy $p_{c}=\frac{1}{T}$ for $<k>\to\infty$, we adopt the form
\begin{equation}
p_{c}\sim\frac{1}{T}+\frac{1}{<k>T}.
\end{equation}

Next, consider the effect of rewiring on the epidemic. At each time step, because of the cut of S-I links, the average degree of the infected decreases, which leads to the rise of the epidemic threshold. Equation (13) can be rewritten as
\begin{equation}
p_{c}=\frac{1}{T}+\frac{1}{T\kappa},
\end{equation}
where $\kappa$ represents the average degree of the infected having lost a fixed ratio $\frac{\omega T}{a^s}$ of its links. For an exponential expression $k(t)=<k>e^{{-\omega Tt}/a^s}$[13], we get
\begin{equation}
\kappa=\frac{\int_{0}^{T}<k>e^{-\omega Tt/a^s}dt}{T}.
\end{equation}
Substituting equation (15) into equation (14), we obtain the epidemic threshold in the present model
\begin{equation}
p_{c}=\frac{1}{T}+\frac{\omega T}{<k>a^s(1-e^{-\omega T^2/a^s})}.
\end{equation}

\begin{figure}
\includegraphics[width=3cm]{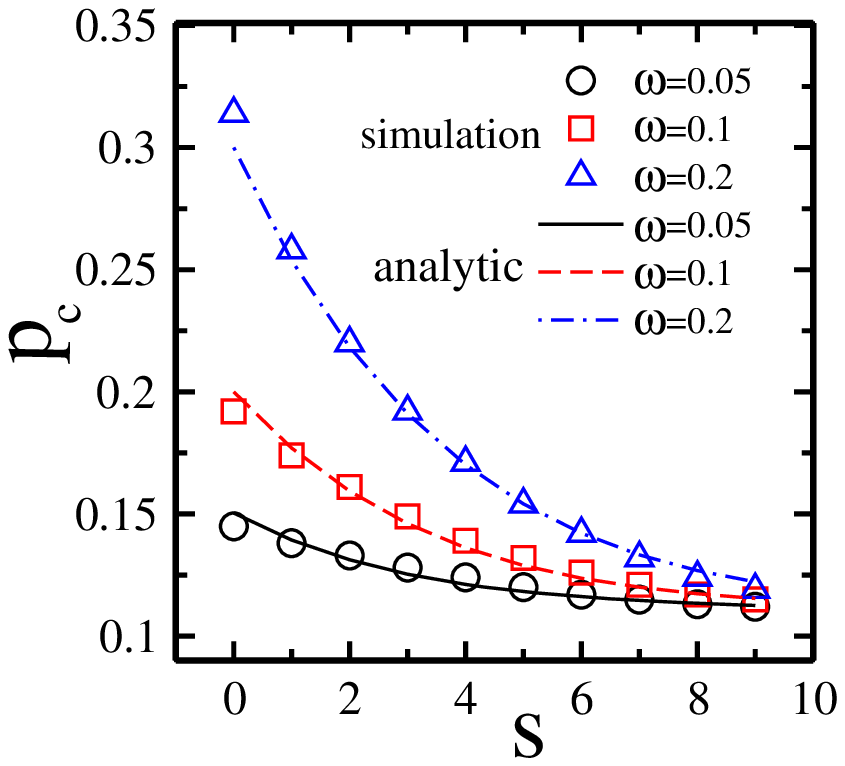}
\caption{\label{fig:epsart}The epidemic threshold as a function of s for $\omega=0.05$(circles), 0.1(squares), and 0.2(triangles) in simulation and analytical calculation(lines).}
\end{figure}

In Fig.7 we give the epidemic threshold as a function of s for $\omega=0.05$, 0.1, and 0.2, the analytical calculations are in accordance with the simulation data.

\section{\label{sec:level5}Summary}
The present work has offered us a clear picture of the epidemic spread pattern in the dynamically structured populations. It is found that the effectiveness of refraining from being infected by the infectious disease is constrained by the epidemic information known to the public. The erroneous reconnections posed by the lagged epidemic information would result in the prolonged epidemic which in turn exacerbates the heterogeneity of the evolved social networks. A power-law relation between the standard deviation of the degree distribution and the duration time before the epidemic becomes stable is found. Depending upon the mean field theory, we have analytically calculate the degree distribution and the epidemic threshold in which the time scale parameter is included.

The understanding of the time scale effect presented here could help us accurately estimate the realistic risks caused by the delay in taking measures and develop a variety of disease control methods to avoid the epidemic outbreaks. In the future, the generalized epidemic model should be further studied in the human mobility circumstances and the epidemic spread pattern in the community structures is one of the favorite interests of us.

\begin{acknowledgments}
This work is the research fruit of Social Science Foundation of Zhejiang Province (Grant No.10CGGL14YB), General Projects of Social Science Research Fund of Ministry of Education of China and the National Natural Science Foundation of China (Grant Nos. 10905023, 10805025, and 10774080).
\end{acknowledgments}



\nocite{*}

\begin{thebibliography}{99}

\bibitem{tang} M. Tang, L. Liu, and Z. H. Liu,
Phys. Rev. E \textbf{79}, 016108 (2009).

\bibitem{parshani} R. Parshani, S. Carmi, and S. Havlin, Phys. Rev. Lett. \textbf{104}, 258701 (2010).

\bibitem{moore} C. Moore and M. E. J. Newman, Phys. Rev. E \textbf{61}, 5678 (2000).

\bibitem{zhou} T. Zhou, G. Yan, and B. H. Wang, Phys. Rev. E \textbf{71}, 046141 (2005).

\bibitem{pacheco} J. M. Pacheco, A. Traulsen, and M. A. Nowak, Phys. Rev. Lett. \textbf{97}, 258103 (2006).

\bibitem{grabowski} A. Grabowski and R. A. Kosinski, Phys. Rev. E \textbf{70}, 031908 (2004).

\bibitem{fefferman} N. H. Fefferman and K. L. Ng, Phys. Rev. E \textbf{76}, 031919 (2007).

\bibitem{kuperman} M. Kuperman and G. Abramson, Phys. Rev. Lett. \textbf{86}, 2909 (2001).

\bibitem{barthelemy} M. Barthelemy, A. Barrat, R. Pastor-Satorras, and A. Vespignani, Phys. Rev. Lett. \textbf{92}, 178701 (2004).

\bibitem{zheng} D. F. Zheng, P. M. Hui, S. Trimper, and B. Zheng, Physica A \textbf{352}, 659 (2005).

\bibitem{colizza} V. Colizza and A. Vespignani, Phys. Rev. Lett. \textbf{99}, 148701 (2007).

\bibitem{watts} D. J. Watts, R. Muhamad, D. C. Medina, and P. S. Dodds, Proc. Natl. Acad. Sci. U. S. A.\textbf{102}, 11157 (2005).

\bibitem{gross} T. Gross, Carlos J. Dommar D'Lima, and B. Blasius, Phys. Rev. Lett. \textbf{96}, 208701 (2006).

\bibitem{shaw} L. B. Shaw, I. B. Schwartz, Phys. Rev. E \textbf{77}, 066101 (2008).

\bibitem{son} S. W. Son, B. J. Kim, H. Hong, H. Jeong, Phys. Rev. Lett. \textbf{103}, 228702 (2009).

\bibitem{volz} E. Volz and L. A. Meyers, J. R. Soc. Interface \textbf{6}, 233 (2009).

\bibitem{graser} O. Graser, C. Xu, and P. M. Hui, Europhys. Lett. \textbf{87}, 38003 (2009).

\bibitem{buscarino} A. Buscarino, L. Fortuna, M. Frasca, and V. Latora, Europhys. Lett. \textbf{82}, 38002 (2008).

\bibitem{zhao} Z. Zhao, J. P. Calderon, C. Xu, G. Zhao, D. Fenn, D. Sornette, R. Crane, P. M. Hui, and N. F. Johnson, Phys. Rev. E \textbf{81}, 056107 (2010).

\bibitem{ni} S. J. Ni and W. G. Weng, Phys. Rev. E \textbf{79}, 016111 (2009).

\bibitem{roca} C. P. Roca, J. A. Cuesta, and A. Sanchez, Phys. Rev. Lett. 97, \textbf{158701} (2006).

\bibitem{zhong1} L. X. Zhong, D. F. Zheng, B. Zheng, C. Xu, and P. M. Hui, Europhys. Lett. \textbf{76}, 724 (2006).

\bibitem{mosetti} G. Mosetti, D. Challet, and Y. C. Zhang, Physica A \textbf{365}, 529 (2006).

\bibitem{funk} S. Funk, E. Gilad, C. Watkins, and V. A. A. Jansen, Proc. Natl. Acad. Sci. U. S. A \textbf{106}, 6872 (2009).

\bibitem{bagnoli} F. Bagnoli, P. Lio, L. Sguanci, Phys. Rev. E \textbf{76}, 061904 (2007).

\bibitem{khan} Q. J. A. Khan and E. V. Krishnan, Applications of Mathematics \textbf{48}, 193 (2003).

\bibitem{xu} X. J. Xu, H. O. Peng, X. M. Wang, and Y. H. Wang, Physica A \textbf{367}, 525 (2006).

\bibitem{dodds1} P. S. Dodds and D. J. Watts, Phys. Rev. Lett. \textbf{92}, 218701 (2004).

\bibitem{barabasi} A. L. Barabasi and R. Albert, Science \textbf{286}, 509 (1999).

\bibitem{newman} M. E. J. Newman, SIAM Review \textbf{45}, 167 (2003).

\bibitem{zhong2} L. X. Zhong, F. Ren, T. Qiu, J. R. Xu, B. H. Chen, C. F. Liu, Physica A \textbf{389}, 2557 (2010).

\bibitem{dodds2} P. S. Dodds, D. J. Watts, J. Theor. Biol. \textbf{232}, 587 (2005).

\end{thebibliography}
\end{document}